\documentclass[letterpaper, 10 pt, conference]{ieeeconf}  

\IEEEoverridecommandlockouts                              

\usepackage[sort,compress]{cite}
\usepackage{graphicx,times}
\usepackage{amsmath,amssymb}
\usepackage{listings}

\title{\LARGE \bf
Prediction of Influenza A Virus Infections in Humans using an Artificial Neural Network Learning Approach
}

\author{Charalambos Chrysostomou$^{1*}$, Harris Partaourides$^{2}$ and Huseyin Seker$^{3}$
\thanks{$^{1}$Computation-based Science and Technology Research Center, The Cyprus Institute, 20 Konstantinou Kavafi Street, 2121, Aglantzia, Nicosia, Cyprus}%
\thanks{$^{2}$Department of Electrical Engineering, Computer Engineering and Informatics, Cyprus University of Technology, 30 Archbishop Kyprianou Str., 3036 Limassol, Cyprus}%
\thanks{$^{3}$Department of Computer and Information Sciences, Faculty of Engineering and Environment, The University of Northumbria at Newcastle, NE1 8ST, Newcastle-upon-Tyne, The United Kingdom}%
\thanks{c.chrysostomou@cyi.ac.cy, c.partaourides@cut.ac.cy, huseyin.seker@northumbria.ac.uk}%
\thanks{*Corresponding Author}%
}

\begin{document}

\maketitle
\thispagestyle{empty}
\pagestyle{empty}

\begin{abstract}

The Influenza type A virus can be considered as one of the most severe viruses that can infect multiple species with often fatal consequences to the hosts. The Haemagglutinin (HA) gene of the virus has the potential to be a target for antiviral drug development realised through accurate identification of its sub-types and possible the targeted hosts. In this paper, to accurately predict if an Influenza type A virus has the capability to infect human hosts, by using only the HA gene, is therefore developed and tested. The predictive model follows three main steps; (i) decoding the protein sequences into numerical signals using EIIP amino acid scale, (ii) analysing these sequences by using Discrete Fourier Transform (DFT) and extracting DFT-based features, (iii) using a predictive model, based on Artificial Neural Networks and using the features generated by DFT. 

In this analysis, from the Influenza Research Database, 30724, 18236 and 8157 HA protein sequences were collected for Human, Avian and Swine respectively. Given this set of the proteins, the proposed method yielded 97.36\% ($\pm$ 0.04\%), 97.26\% ($\pm$ 0.26\%), 0.978 ($\pm$ 0.004), 0.963 ($\pm$ 0.005) and 0.945 ($\pm$ 0.005) for the training accuracy validation accuracy, precision, recall and Mathews Correlation Coefficient (MCC) respectively, based on a 10-fold cross-validation.  The classification model generated by using one of the largest dataset, if not the largest, yields promising results that could lead to early detection of such species and help develop  precautionary measurements for possible human infections.

\end{abstract}

\begin{keywords}
Artificial Neural Network, Amino Acid Indices, Discrete Fourier Transform (DFT), Hemagglutinin (HA) Protein
\end{keywords}

\section{INTRODUCTION}

The Influenza type A virus can be considered one of the most severe virus that can infect both mammals and birds. The genome of the Influenza virus is composed of eight segments that can encode more than 11 proteins \cite{webster1992evolution}. One of the most important proteins is the Haemagglutinin (HA), which is an essential glycoprotein and a principal surface antigen which is responsible for attaching the virions to hosts, deciding the pathogenicity and virulence \cite{webster1992evolution}.  Until now, 18 distinct Influenza A HA subtypes have been identified \cite{fouchier2005characterization, wu2014bat}. 

The Influenza type A virus continually evolves due to the high mutation rate and the constant changes to its genome.  This constant adaptation usually makes any new strain of virus more pathogenic than the previous. Furthermore, these mutations also provide the virus with the ability to cross the species barrier and may also affect the binding pattern of a virus, with catastrophic consequences to the concerned species \cite{nunthaboot2010evolution}.

In the literature, previous efforts and analysis have been performed to analyse and characterise the phylogenetic diversity, and discover mechanisms that define the severity and distribution of influenza type A virus \cite{chen2009panorama, liu2009panorama, shi2010complete}. Additionally, as the authors concluded, classification and characterisation of all the sequences with the proposed methods, was difficult \cite{chen2009panorama, liu2009panorama, shi2010complete}, thus a more advanced method is needed. Computational studies exist that tries to characterise and analyse the Influenza type A with promising results \cite{rehman2016characterization}. In the proposed method a computational, Artificial Neural Network (ANN) learning based approach is created to predict if a particular virus has the capability to infect humans, by only analysing the HA protein sequence.

The paper is organised as follows: Section \ref{section:methodsandmaterials} presents the methods and materials developed and used, while Section \ref{section:results} presents the results obtained. Finally, concluding remarks are outlined in Section \ref{section:conclusions}.

\section{METHODS AND MATERIAL}
\label{section:methodsandmaterials}

\subsection{Influenza A Hemagglutinin Proteins Data Set}

\begin{figure*}[ht]
    \centering
    \includegraphics[width=1.0\textwidth, height=0.40\textheight]{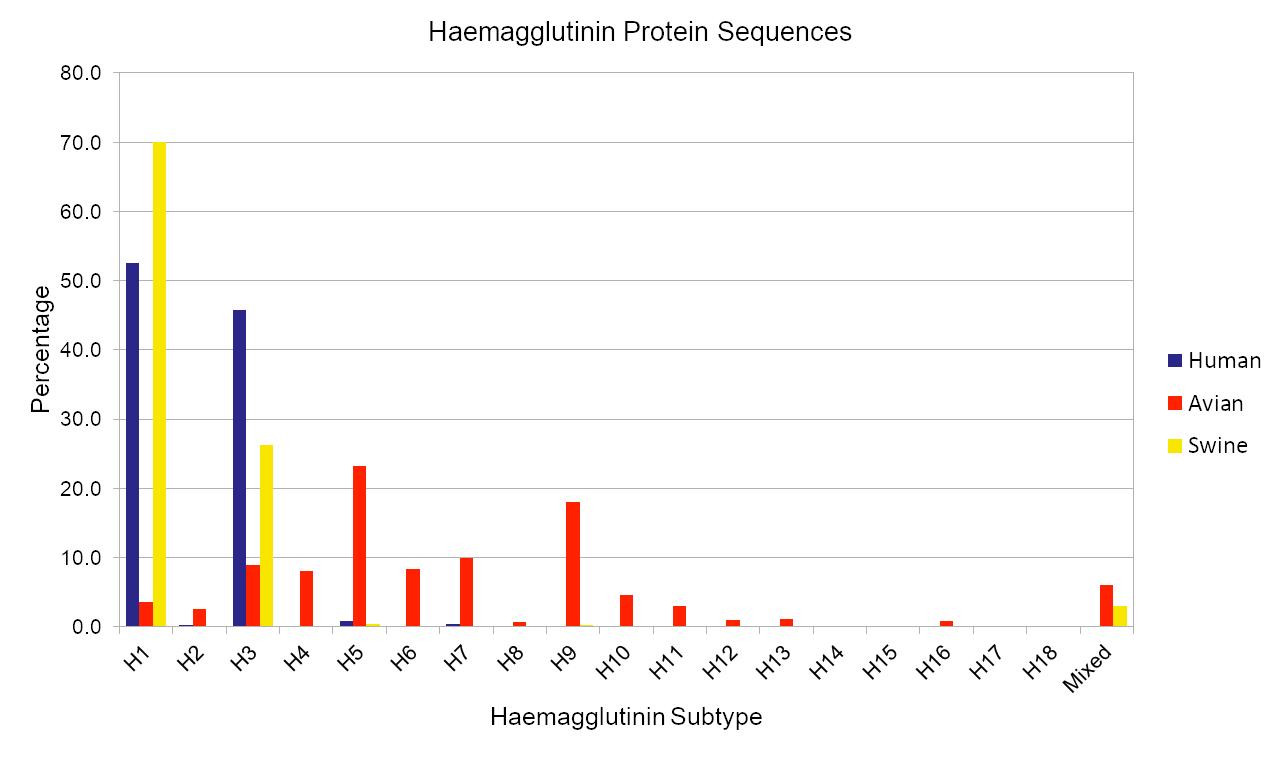}
    \caption{HA Protein Sequences}
    \label{HA_protein_sequences_figure}
\end{figure*}

For the proposed analysis 57117 HA Influenza type A protein sequences are collected from the Influenza Research Database \cite{squires2012influenza}, for three species, Human, Avian and Swine. More specifically, as Table \ref{number_of_HA_proteins} shows 30724, 18236 and 8157 HA protein sequences were collected for Human, Avian and Swine respectively. Furthermore, Table \ref{number_of_HA_proteins} shows the specific number of sequences for each class of HA 1-18. Finally, figure \ref{HA_protein_sequences_figure} illustrates the percentage of HA proteins per class and per species. For the analysis, classification of the HA protein sequences, based on the ability of the virus to infect human hosts, the data were separated into two groups. The first group contained all the sequences from HA 1-18 for the viruses that have the ability to infect the Humans hosts, and the second group with the sequences that have the potential to infect the Avian and Swine hosts. For the first and second groups, the total number of 30724 and 26393 HA protein sequences were used respectively.

\begin{table}
\caption{Number of HA Protein Sequences}
\centering
\begin{tabular}{c c c c}
\hline
 HA Subtype & Human & Avian & Swine\\
 \hline
H1 & 16145 & 650 & 5714\\
H2 & 96 & 462 & 2\\
H3 & 14055 & 1621 & 2138\\
H4 & 0 & 1478 & 5\\
H5 & 269 & 4242 & 32\\
H6 & 0 & 1529 & 2\\
H7 & 104 & 1809 & 1\\
H8 & 0 & 121 & 0\\
H9 & 13 & 3281 & 20\\
H10 & 4 & 834 & 1\\
H11 & 0 & 547 & 0\\
H12 & 0 & 178 & 0\\
H13 & 0 & 200 & 0\\
H14 & 0 & 17 & 0\\
H15 & 0 & 13 & 0\\
H16 & 0 & 150 & 0\\
H17 & 0 & 0 & 0\\
H18 & 0 & 0 & 0\\
Mixed & 38 & 1104 & 242\\
\hline
Total & 30724 & 18236 & 8157\\
\hline
\end{tabular}
\label{number_of_HA_proteins}
\end{table}

\subsection{Data conversion and Normalisation}

In this paper digital signal processing techniques are used to extract information that can be directly used to characterise the HA proteins. In the literature, various methods used signal processing in bioinformatics for analysing and characterising protein sequences \cite{chrysostomou2010complex, carmona2013fuzzy, chrysostomou2011effects, chrysostomou2015cisaps, chrysostomou2016structural} such as Complex resonant recognition model in analysing influenza a virus subtype protein sequences \cite{chrysostomou2010complex}, CISAPS: Complex informational spectrum for the analysis of protein sequences \cite{chrysostomou2015cisaps} and Structural classification of protein sequences based on signal processing and support vector machines \cite{chrysostomou2016structural}. Furthermore, previous studies \cite{flu_characterization_2009} where signal processing was used to analyse influenza A HA proteins aimed to identify new therapeutic targets for drug development by better understanding the interaction of the influenza virus and its receptors.

For the proposed analysis, signal processing methods are used, and more specifically Discrete Fourier Transform (DFT), as shown in equations 1-3. The analysis was performed directly to absolute spectrum. Before applying DFT to the HA protein sequences, Electron-ion interaction potential (EIIP) \cite{v_is_1985, gopalakrishnan_computational_2004} amino acid index, was used to convert alphanumerical sequences. The complete list of the EIIP amino acid index can be found in Table \ref{eiip_values}. 

\vspace{3 mm}

\noindent Discrete Fourier Transform (DFT)

\begin{equation}
X(n) = \sum^{N-1}_{m=0} {x(m)e^{-j(2\pi /N)nm}}  \:\:\:\:\:  n = 0,1,...,N-1
\end{equation}
\vspace{3 mm}

\noindent where $X(n)$ are the DFT coefficients, N is the total number of points in the series and $x(m)$ is the $m$th member of the numerical series. As the DFT coefficients contain two mirror parts, only the $(N/2)$ points of the series will be used.

\noindent The output of DFT is a complex sequence and can be formulated as

\begin{equation}
X(n) = (R(n) + jI(n)), \:\:\:\:  n = 0,1,...,(N-1)/2
\end{equation}

\vspace{3 mm}
\noindent where $R(n)$ and $I(n)$ are the Real and Imaginary parts of the sequence, respectively.
\noindent The absolute spectrum ($S_{(n)}$) can be formulated as 

\begin{equation}
S_{(n)} = X(n)X^{*}(n) = \left|X(n)\right|^{2}, \:\:\:\:  n = 0,1,...,(N-1)/2
\end{equation}

\vspace{3 mm}
\noindent where $X(n)$ are the DFT coefficients of the series $x(n)$, $X*(n)$ are the complex conjugates.

\vspace{5 mm}

The coefficients from the absolute spectrum will be used as a feature set to represent the characteristics of different classes of proteins' secondary structure. The HA influenza A virus proteins sequences have different lengths, and zero-padding was used to extend all the protein sequences to $N = 1024$ before applying DFT. After DFT is applied the output of the absolute spectrum includes 513 features. These features are used as input to the ANN model.  

\begin{table}[ht]
\caption{EIIP Values}
\setlength{\tabcolsep}{6pt}
\centering 
\begin{tabular}{l c} 
\hline 
Amino acid & EIIP Values \\
\hline
Leucine & 0.0000\\
Isoleucine & 0.0000\\ 
Asparagine & 0.0036\\ 
Glycine & 0.0050\\ 
Glutamic acid & 0.0057\\ 
Valine & 0.0058\\ 
Proline & 0.0198\\ 
Histidine & 0.0242\\ 
Lysine & 0.0371\\ 
Alanine & 0.0373\\ 
Tyrosine & 0.0516\\
Tryptophan & 0.0548\\
Glutamine & 0.0761\\
Methionine & 0.0823\\
Serine & 0.0829\\
Cysteine & 0.0829\\
Threonine & 0.0941\\
Phenylalanine & 0.0946\\
Arginine & 0.0959\\
Aspartic acid & 0.1263\\

  \hline 
\end{tabular} 
\label{eiip_values} 
\end{table}

\subsection{Artificial Neural Network - Experimental Evaluation}

Artificial Neural Networks (ANN) \cite{gupta2013artificial} are a computational method, based on an extensive collection of artificial neurons, which mirrors the process a living brain solves problems. Each neuron connects to multiple other neurons, which can enforce or repress the impact on the activation event of the connected neurons. The ANNs are considered as self-trained, rather than explicitly programmed, and employed in research fields where the discovery of features and classification is challenging in traditional classification systems.   
 
For the proposed work, the ANN receives an input of 513 features derived from the influenza type A virus pre-processing and returns the probability of the virus infecting humans. For the proposed work, the binary classification consist of 57117 samples of which 30724 can infect humans. 

\begin{figure}[ht]
    \centering
    \includegraphics[width=0.4\textwidth, height=0.30\textheight]{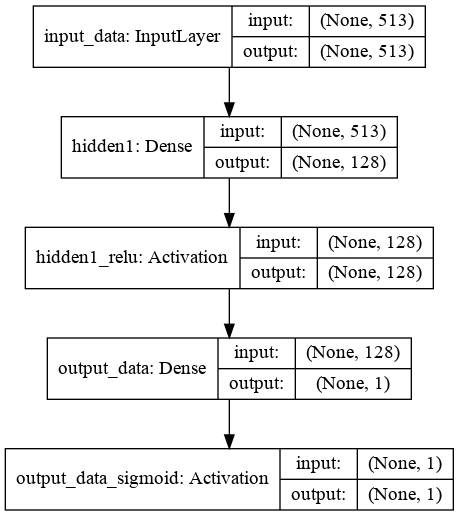}
    \caption{Artificial Neural Network Model}
    \label{NN_figure}
\end{figure}

The network setup consists of a single hidden layer of 128 units, Glorot-style uniform for initialization and rectified linear units for the activation function. In order to train the ANN, the Adam optimizer \cite{DBLP:journals/corr/KingmaB14} was used with mini batch size of 128 for 200 epochs. We use 10 fold cross-validation and show the network performance based on accuracy. The model was implemented by utilising the Tensorflow \cite{abadi2016tensorflow} and Keras \cite{chollet2015keras} libraries. A visual representation of the model can be seen in Figure \ref{NN_figure}.

The high performance of both training and testing shows that for this type of problem more advanced Neural Network models (such as Deep Neural Networks) and regularization techniques are not needed.

\section{RESULTS AND DISCUSSION}
\label{section:results}

In this paper, a classification model is presented, based on Artificial Neural Networks, for the analysis and classification Influenza type A based upon the ability to infect a human host solely by using the HA protein sequence. To ensure that the proposed classification model is accurate and the results can be generalised, 10-fold cross-validation was used. 
The total accuracy of the predictive model with average training accuracy, testing accuracy, precision, recall and MCC of 97.36\% ($\pm$ 0.04\%), 97.26\% ($\pm$ 0.26\%), 0.978 ($\pm$ 0.004), 0.963 ($\pm$ 0.005) and 0.945 ($\pm$ 0.005) for the training accuracy validation accuracy, precision, recall and Mathews Correlation Coefficient (MCC) respectively. As the results show, the proposed model can distinguish HA protein sequences with extremely high accuracy whenever the virus under investigation will have the capability to infect human hosts. Detailed results can be found in Table \ref{Results}.

\begin{table*}[ht]
\caption{Accuracy Results for the Prediction of Influenza A Virus Infections}
\centering
\begin{tabular}{c c c c c c}
\hline

Fold & Training Accuracy & Validation Accuracy & Validation Precision & Validation Recall & Validation MCC\\
\hline

1 & 0.974 & 0.970 & 0.976 & 0.959 & 0.940\\
2 & 0.973 & 0.971 & 0.982 & 0.955 & 0.941\\
3 & 0.973 & 0.970 & 0.973 & 0.962 & 0.940\\
4 & 0.974 & 0.971 & 0.974 & 0.964 & 0.942\\
5 & 0.974 & 0.970 & 0.982 & 0.954 & 0.941\\
6 & 0.973 & 0.975 & 0.981 & 0.966 & 0.950\\
7 & 0.974 & 0.972 & 0.973 & 0.967 & 0.944\\
8 & 0.973 & 0.978 & 0.982 & 0.971 & 0.957\\
9 & 0.974 & 0.975 & 0.980 & 0.966 & 0.950\\
10 & 0.974 & 0.972 & 0.976 & 0.963 & 0.944\\
\hline 
Average & 97.36\% ($\pm$ 0.04\%) & 97.26\% ($\pm$ 0.26\%) & 0.978 ($\pm$ 0.004) & 0.963 ($\pm$ 0.005) & 0.945 ($\pm$ 0.005)\\
\hline
\end{tabular}
\label{Results}
\end{table*}


\section{CONCLUSIONS}
\label{section:conclusions}

The paper presents a highly successful predictive model to identify and differentiate Influenza type A virus, which can and cannot infect Humans, based on the HA gene, which is considered as a highly potential antiviral drug candidate. The classification model was created by utilising one of the largest possible datasets, if not the largest, in order to better generalise the model. The classification model obtained over the 10-fold cross validation yielded the average training accuracy, testing accuracy, precision, recall and MCC of 97.36\% ($\pm$ 0.04\%), 97.26\% ($\pm$ 0.26\%), 0.978 ($\pm$ 0.004), 0.963 ($\pm$ 0.005) and 0.945 ($\pm$ 0.005) respectively. Also, by applying signal processing technique, namely Discrete Fourier Transform, on the protein sequences, it was found that useful spectral characteristic features can be distinguished that are capable of representing the protein groups and thus further enhanced using the Artificial Neural Network based classifier. 

As reported in the literature, there are over 600 amino acid indices, where each represents a unique physicochemical feature of the protein \cite{chrysostomou2015cisaps}, in contrast to the one amino acid index used throughout this study. Future studies are required to identify any potential amino acid indices that are capable of representing, characterising and classify the Influenza type A HA protein sequences. A computational tool that is capable of classifying and distinguishing potentially dangerous viruses that have the capability to infect Human hosts will be critical and required to monitor future outbreaks. Future studies will investigate additional machine learning approaches, to explore the efficiency of the methodology, using signal processing methods to encode protein sequences.






\bibliographystyle{IEEEtran}
\bibliography{papers.bib}

\end{document}